\documentclass[letterpaper, 10pt, conference]{ieeeconf}  

\IEEEoverridecommandlockouts                              

\usepackage{ntheorem}
\usepackage{subfigure, amsmath, rotating}
\usepackage{array, multirow}
\usepackage{amsbsy, amsfonts, graphics, graphicx}
\usepackage{algorithmic}
\usepackage{algorithm}
\usepackage{epstopdf}
\usepackage{mathrsfs}
\usepackage{graphicx}
\usepackage{caption}
\usepackage{amssymb}
\usepackage{amsmath,bm} 

\usepackage{subfiles}

\usepackage[multiple]{footmisc}
\usepackage{mathtools}

\usepackage{tikz}
\usepackage{float}
\usetikzlibrary{calc,patterns,decorations.pathmorphing,decorations.markings}

\newcommand*{\Scale}[2][4]{\scalebox{#1}{$#2$}}%
\newtheorem{assumption}{Assumption}
\newtheorem{theorem}{Theorem}

\newtheorem{remark}{Remark}
\newtheorem{lemma}{Lemma}

\newtheorem{proposition}{Proposition}
\renewcommand{\qedsymbol}{\rule{0.55em}{0.55em}}

\title{\LARGE \bf Sub-Optimal Moving Horizon Estimation in Feedback Control of Linear Constrained Systems
}

\author{Yujia Yang, Chris Manzie, and Ye Pu
\thanks{$^{1}$Y. Yang, C. Manzie, and Y. Pu are with the Department of Electrical and Electronic Engineering, University of Melbourne, Parkville VIC 3010, Australia {\tt\small {yujyang1}@student.unimelb.edu.au, {manziec,ye.pu}@unimelb.edu.au}}}
\begin{document}

\maketitle
\thispagestyle{empty}
\pagestyle{empty}

\begin{abstract}

Moving horizon estimation (MHE) offers benefits relative to other estimation approaches by its ability to explicitly handle constraints, but suffers increased computation cost.
To help enable MHE on platforms with limited computation power, we propose to solve the optimization problem underlying MHE sub-optimally for a fixed number of optimization iterations per time step.
The stability of the closed-loop system is analyzed using the small-gain theorem by considering the closed-loop controlled system, the optimization algorithm dynamics, and the estimation error dynamics as three interconnected subsystems.
By assuming incremental input/output-to-state stability ($\delta$-IOSS) of the system and imposing standard ISS conditions on the controller, we derive conditions on the iteration number such that the interconnected system is input-to-state stable (ISS) w.r.t. the external disturbances.
A simulation using an MHE-MPC estimator-controller pair is used to validate the results.
\end{abstract}

\section{introduction} \label{sec intro}

MHE is an optimization-based method that considers a fixed window of past measurements and the system's constraints in estimating the current state.
Due to the inclusion of the constraints explicitly in the problem formulation, MHE has been shown to produce more accurate state estimates compared to the extended Kalman Filter \cite{mpc_book}.
Assuming detectability of the system, rather than observability, MHE was shown to posses robust global asymptotic stability w.r.t. bounded disturbances and the estimation error converges in case of bounded and vanishing disturbances \cite{muller_nonlinear}.

Although MHE offers the benefit of considering constraints, its application is limited by the computational cost, particularly in systems with fast dynamics or platforms with limited computational resources. To alleviate this issue, \cite{MHE_PE} introduced an auxiliary observer to provide pre-estimation for MHE. However, despite reduced computation time, the iteration number required to solve the MHE problem with stability guarantees cannot be determined offline.
In \cite{muller_sub_mhe}, a feasible candidate solution from an auxiliary observer is improved for a limited but varying amount of iterations to obtain a sub-optimal solution so that the resulting estimate is robustly stable.
The proximity-MHE scheme in \cite{anytime_mhe} performs limited optimization iterations with a proximity regularizing term to improve the prior estimate from an auxiliary observer and guarantees the nominal stability of the MHE.

Other approaches concentrated on the optimization scheme that underlies the MHE problem. For example, \cite{KONG201866} proposed to enforce move blocking on the disturbance sequence in MHE to reduce the associated computation burden, which also guarantees the nominal stability of MHE.
In \cite{RTI}, a real-time iteration scheme is applied to MHE without inequality constraints. Local convergence is guaranteed when a single optimization iteration is performed per time step.
The work \cite{6426428} combined this scheme with automatic code generation to obtain highly efficient source code of MHE algorithms.
For noise-free systems, \cite{RTI_multiple} solves the MHE problem for single or multiple iterations with gradient-based, conjugate gradient-based, and Newton methods and achieves local stability.

Compared to the aforementioned works, we study the stability of the closed-loop with a sub-optimal MHE and a feedback control law.
Earlier studies often treated MHE and the feedback controller as separate modules, with MHE providing estimates with bounded error \cite{MHE_MPC}, and the controller designed to ensure stability.
Instead, we aim to jointly determine conditions that guarantee stability of both MHE and the controlled system.
To achieve this, we employ an stability analysis framework from the sub-optimal model predictive control (MPC) literature \cite{dominic1},\cite{zanelli2021lyapunov},\cite{our_lcss}.
Therein, the closed-loop system was formulated as an interconnection of a controlled system and an optimization algorithm dynamics.

In this paper, we propose a sub-optimal MHE scheme where, at every time step, the MHE problem is warm-started with the previous solution and then solved by an optimization algorithm with a fixed number of iterations. Then, the resulting sub-optimal estimate is used for feedback control of a linear system with state and input constraints. 

Our main contribution lies in the stability analysis, which follows a similar approach as \cite{dominic1},
\cite{zanelli2021lyapunov}, and \cite{our_lcss}.
We first characterize the interaction between the closed-loop controlled system, the sub-optimality error dynamics (of the optimization algorithm used for solving the MHE problem), and the state estimation error dynamics as three interconnected subsystems.
Then, assuming the controller is robustly stabilizing, the small-gain theorem is used to derive conditions on the optimization iteration number for guaranteeing the interconnected system is input-to-state stable (ISS) w.r.t to the external disturbances.
\vspace{4pt}




\noindent \textup{\textbf{Notations:}} 
Let $\mathbb{S}_{\succ 0}$ be the set of positive definite matrices.
Let $\mathbf{I}^{n}$ be the identity matrix of size $n$.
Let $\mathbf{0}^{m \times n}$ be the zero matrix of size $m \times n$.
For a vector $x \in \mathbb{R}^{n_x}$ and a matrix $U \in \mathbb{S}^{n_x \times n_x}_{\succ 0}$,
let $\|x\|$ and $\|x\|_U$ denote the $l_2$-norm and the weighted $l_2$-norm of $x$, respectively.
Consider square matrices $U$ and $V$. 
Let $\| U \|$ denote the spectral norm.
Let $\overline{\lambda}_U$ and $\underline{\lambda}_U$ denote the largest and smallest eigenvalues of $U$, respectively.
Let $\Lambda^U_{V} := \overline{\lambda}(U) / \underline{\lambda}(V)$.
Let $\mathbb{I}_{[a, b]}$ denote the set of integers in $[a, b] \in \mathbb{R}$.
For a variable $v_t \in \mathbb{R}^{n_v}$ and time steps $a, b \in \mathbb{I}_{[a, b]}$, let $\mathbf{v}_{[a,b]} := \{v_a,\cdots,v_b\}$ and $\|\mathbf{v}_{[a,b]}\| := \sup_{t \in \mathbb{I}_{[a, b]}} \| v_t\|$.
A continuous function $\gamma: \mathbb{R}_+ \rightarrow \mathbb{R}_+$ is of class $\mathcal{K}$ if it is strictly increasing and $\gamma(0) = 0$. If it is also unbounded, then it is of class $\mathcal{K}_{\infty}$. If $\gamma$ is strictly decreasing and $\gamma(s) \rightarrow 0$ as $s \rightarrow 0$, then it is of class $\mathcal{L}$.
A continuous function $\beta: \mathbb{R}_+ \times \mathbb{R}_+ \rightarrow \mathbb{R}_+$ is of class $\mathcal{KL}$ if $\beta(\cdot,s) \in \mathcal{K}$ for each fixed $s$ and $\beta(r,\cdot) \in \mathcal{L}$ for each fixed $r$.


\section{Controller and MHE Formulation} \label{sec tight}

\subsection{Dynamic System with State Feedback Controller}
Consider a system with linear time-invariant dynamics 
\begin{equation}
\begin{aligned} \label{controlled system}
x_{t+1} &= A x_t + B u_t + w^1_{t}, \\
y_t &=  C x_t + w^2_{t},
\end{aligned}
\end{equation}
with state $x_t \in \mathcal{X} \subset \mathbb{R}^{n_x}$, input $u_t \in \mathcal{U} \subset \mathbb{R}^{n_u}$, output measurement $y_t \in \mathcal{Y} \subset \mathbb{R}^{n_y}$, external disturbance $w^1_{t} \in \mathcal{W}_1 \subset \mathbb{R}^{n_{x}}$, and measurement noise $w^2_{t} \in \mathcal{W}_2 \subset \mathbb{R}^{n_{y}}$.
Let $w_t := [{w^1_{t}}^\top, {w^2_{t}}^\top]^\top \in \mathcal{W} \subset \mathbb{R}^{n_{x+y}}$ be the augmented disturbance.
Let $\mathcal{Z} := \mathcal{X} \times \mathcal{U} \times \mathcal{Y} \times \mathcal{W}$ be the Cartesian product of the constraint sets.

\begin{assumption} \label{constraints assumption}
    $\mathcal{Z}$ is convex and contains the origin.
\end{assumption}

\begin{assumption} \label{delta-IOSS lyapunov}
Consider system \eqref{controlled system}.
 There exist $P, Q, R \in \mathbb{S}_{\succ 0}$ and $\eta \in [0,1)$ that satisfy\
\begin{align} \label{ioss condition}
  & \left(\begin{array}{cc}
A^{\top} P A-\eta P-C^{\top} R C & A^{\top} P \Bar{B} -C^{\top} R \Bar{D} \\
\Bar{B}^\top P A- \Bar{D}^\top R C &  \Bar{B}^\top P \Bar{B} -Q-\Bar{D}^\top R \Bar{D}
\end{array}\right) \preceq 0, \nonumber \\
& \Bar{B} = [\textbf{\textup{I}}^{n_x}, \textbf{\textup{0}}^{n_x \times n_y}], \quad \Bar{D} = [\textbf{\textup{0}}^{n_y \times n_x}, \textbf{\textup{I}}^{n_y}, ].
\end{align}
\end{assumption}
From Corollary 3 of \cite{muller_lyapu}, we know Assumption \ref{delta-IOSS lyapunov} implies system \eqref{controlled system} admits a $\delta$-IOSS Lyapunov function and is detectable.
Specifically, for $(x, u, y, w),(x', u, y', w') \in \mathcal{Z}$, where $y = C x + w^{2}$ and $y' = C x'+ {w}^{2} {'}$, the function
\begin{align} \label{ioss function definition}
    W_\delta(x, x')=\|x-x'\|_{P}^2
\end{align}
is a $\delta$-IOSS Lyapunov function for system \eqref{controlled system}, satisfying 
\begin{align} \label{ioss function}
& W_\delta( A x + B u + w^1 ,  A x' + B u + w^{1} {'} ) \nonumber \\
& \quad \leq \eta W_\delta(x, x')+\|w-w'\|_Q^2+\|y-y'\|_R^2.
\end{align}

Consider the system \eqref{controlled system} with a state feedback controller $u_t := \pi(\hat{x}_{t}): \mathcal{X} \rightarrow \mathcal{U}$ satisfying Assumptions \ref{lipschitz control} and \ref{system ISS},
\begin{align} \label{re written system}
    x_{t+1} &= A x_t + B \pi(\hat{x}_{t}) + w^1_{t},
\end{align}
where $\hat{x}_t \in \mathcal{X}$ is a state estimate with estimation error $e_t := \hat{x}_t - x_t$.

\begin{assumption} \label{lipschitz control}
 There exists a positive constant $L_{\pi}$ such that, for any $x, x' \in \mathcal{X}$, $\pi(\cdot)$ satisfies 
    \begin{align} \label{controller lip constant}
       \|\pi(x) - \pi(x')\| \leq L_\pi \|x -x'\|.
    \end{align}
\end{assumption}

\begin{assumption} \label{system ISS}
The closed-loop controlled system in \eqref{re written system} is input-to-state stable (ISS):
Given an initial state $x_t \in \mathcal{X}$, an input sequence $\mathbf{u}_{[t,t+g]} \in \mathcal{U} \times \cdots \times \mathcal{U}$ generated from applying $\pi(\cdot)$, an estimation error sequence $\mathbf{e}_{[t,t+g]} \in \mathbb{R}^{n_x} \times \cdots \times \mathbb{R}^{n_x}$, and a disturbance sequence $\mathbf{w}_{[t,t+g]} \in \mathcal{W} \times \cdots \times \mathcal{W}$,
there exist $\beta_1 \in \mathcal{KL}$, and $\gamma_{1,3}, \gamma^{w}_{1} \in \mathcal{K}$ such that, for all $g \geq 0$, the resulting state $x_{t+g} \in \mathcal{X}$ and satisfies  
\begin{align}  \label{sys1 trajectory function}
   \|x_{t+g}\| \leq & \beta_1( \| x_{t} \|, g) + \gamma_{1,3}( \|{\mathbf{e}}_{[t,t+g]}\|) + \gamma^{w}_{1}( \|\mathbf{w}_{[t,t+g]}\|).
\end{align}
\end{assumption}


\subsection{Sub-Optimal Moving Horizon Estimation}
 
At time step $t$, we obtain the state estimate $\hat{x}_t$ by solving a MHE problem based on a prior estimation $x^{\textup{prior}}_{t-M_t}$, past inputs $\mathbf{u}_{[t-M_t,t-1]}$, and past output measurements $\mathbf{y}_{[t-M_t,t-1]}$, with estimation horizon $M_t := \min(M,t)$, $M \in \mathbb{I}_{\geq 0}$.
The MHE problem $\mathbb{P}_t(x^{\textup{prior}}_{t-M_t},{\mathbf{u}}_t,{\mathbf{y}}_t)$ is formulated as
\begin{subequations} \label{MHE formulation}
 \begin{align}
 ( \hat{\mathbf{x}}^*_t, & \hat{\mathbf{w}}^*_t, \hat{\mathbf{y}}^*_t ) = \underset{ \hat{\mathbf{x}}_t, \hat{\mathbf{w}}_t, \hat{\mathbf{y}}_t}{\operatorname{argmin}}  \; V_{\operatorname{MHE}} ( \hat{x}_{t-M_t \mid t}, \hat{\mathbf{w}}_t, \hat{\mathbf{y}}_t ) \\
 \text{s.t. } &\hat{x}_{i+1 \mid t}= A \hat{x}_{i \mid t} + B u_{i} + \hat{w}^1_{i \mid t},\;\;\, i \in \mathbb{I}_{\left[t-M_t, t-1\right]}, \label{sys constraint 1}\\
& \hat{y}_{i \mid t}= C \hat{x}_{i \mid t} + \hat{w}^2_{i \mid t}, \;\;\;\;\;\;\;\;\;\;\;\;\;\;\;\;\; i \in \mathbb{I}_{\left[t-M_t, t-1\right]}, \label{sys constraint 2} \\
& \hat{w}_{i \mid t} \in \mathcal{W}, \; \hat{y}_{i \mid t} \in \mathcal{Y}, \;\;\;\;\;\;\;\;\;\;\;\;\;\;\;\,\; i \in \mathbb{I}_{\left[t-M_t, t-1\right]}, \\
& \hat{x}_{i \mid t} \in \mathcal{X}, \;\;\;\;\; \;\;\;\;\;\;\;\; \;\;\;\;\;\;\;\;\;\;\;\;\;\;\;\;\;\;\;\, i \in \mathbb{I}_{\left[t-M_t, t\right]},
\end{align}   
\end{subequations}
where the cost is defined as
\begin{align} \label{MHE cost}
& V_{\operatorname{MHE}} ( \hat{x}_{t-{M_t} \mid t}, \hat{\mathbf{w}}_t, \hat{\mathbf{y}}_t ) := 2\eta^{M_t}W_\delta(\hat{x}_{t-{M_t} \mid t}, x^{\textup{prior}}_{t-{M_t}}) \nonumber \\
& \;\; + \sum_{i=1}^{M_t} \eta^{i-1}\left(2\left\|\hat{w}_{t-i \mid t}\right\|_Q^2+\left\|\hat{y}_{t-i \mid t}-y_{t-i}\right\|_R^2\right),
\end{align}
with $\eta$, $P$, $Q$, and $R$ satisfying \eqref{ioss condition}.
The decision variables 
$\hat{\mathbf{x}}_t := \{\hat{x}_{t-M_t \mid t},\cdots,\hat{x}_{t \mid t}\}$, 
$\hat{\mathbf{w}}_t := \{\hat{w}_{t-M_t \mid t},\cdots,\hat{w}_{t-1 \mid t}\}$,
and $\hat{\mathbf{y}}_t := \{\hat{y}_{t-M_t \mid t},\cdots,\hat{y}_{t-1 \mid t}\}$ denote the estimated states, augmented disturbances, and measurements, respectively.
The cost functions \eqref{MHE cost} can be reformulated as
\begin{align} \label{MHE cost condensed}
 &   V_{\operatorname{MHE}}(\hat{x}_{t-M_t \mid t}, \hat{\mathbf{w}}_t, \hat{\mathbf{y}}_t ) := \| z_t - \Tilde{z}_t \|^2_{H_t},
\end{align}
where 
\begin{equation} \label{compact definitions}
   \begin{aligned}
   z_t := & [{\hat{x}_{t-{M_t} \mid t}}^\top, {\hat{{w}}_{t-M_t \mid t}}^\top, {\hat{{y}}_{t-M_t \mid t}}^\top, \cdots, {\hat{{w}}_{t-1 \mid t}}^\top, \hat{y}_{t-1 \mid t}^\top]^\top, \\
    \Tilde{z}_t := & [x^{\textup{prior} \top}_{t-{M_t}}, {\mathbf{0}^{n_w}}^\top, y_{t-M_t}^\top, \cdots, {\mathbf{0}^{n_w}}^\top, y_{t-1}^\top]^\top, \\
    H_t := & \operatorname{blkdiag}(2\eta^{M_t} P, 2 \eta^{M_t-1} Q, \eta^{M_t-1} R, \cdots, 2 Q, R).
\end{aligned} 
\end{equation}
Given $\mathbf{u}_t$, the state sequence $\hat{\mathbf{x}}_t$ can be constructed from $z_t$.
Let $(\hat{\mathbf{x}}^*_t,\hat{\mathbf{y}}^*_t,\hat{\mathbf{w}}^*_t)$ and $z^*_t$ denote the optimal solution to \eqref{MHE formulation}, considering the formulations in (9) and (10), respectively.
To solve $\mathbb{P}_t(x^{\textup{prior}}_{t-M_t},{\mathbf{u}}_t,{\mathbf{y}}_t)$, we consider optimization algorithms that satisfies Assumption \ref{optimizer condition}. 

\begin{assumption}\label{optimizer condition}
$\mathbb{P}_t(x^{\textup{prior}}_{t-M_t},{\mathbf{u}}_t,{\mathbf{y}}_t)$ is solved by an optimization algorithm whose iteration can be described by a nonlinear mapping $z^{k+1}_t = \Phi(z^{k}_t,\Tilde{z}_t)$, where $k \geq 0$ is the iteration number.
Furthermore, given an initial solution $z^0_{t}$, the $K_{\textup{th}}$-iteration solution $z^K_{t}$ obtained from applying $\Phi(\cdot)$ for $K$ times is feasible, i.e., satisfying (8b)-(8e), and satisfies
\begin{align} \label{complexity bound}
   \| z^K_{t} - z^*_{t} \| \leq \phi(K) \| z^{0}_{t} - z^*_{t} \|,
\end{align}
where $\phi(K) \in (0,1) \forall K > 0$ and $\phi \in \mathcal{L}$.
\end{assumption}
Let $(\hat{\mathbf{x}}^K_t,\hat{\mathbf{y}}^K_t,\hat{\mathbf{w}}^K_t)$ and $z^K_t$ denote the sub-optimal solution to \eqref{MHE formulation} and define the sub-optimality error as $\epsilon_t := \| z^K_{t} - z^*_{t} \|$.

\section{Sub-Optimal MHE-Based Feedback Control}
In this section, we introduce a sub-optimal MHE scheme.
We characterize the closed-loop system controlled with the proposed scheme as three interconnected subsystems and show each subsystem is ISS.
Lastly, we derive conditions on the optimization iteration number that guarantee the interconnected system is ISS w.r.t. the augmented disturbance ${w}_t$, through the small-gain theorem.
We present the proofs of Propositions 1-3 in the Appendix.

\subsection{The Sub-Optimal MHE Scheme}
Alg. 1 introduces the proposed sub-optimal MHE scheme, employing a warm-start strategy.
When $t < M$, the formulation in \eqref{MHE formulation} represents the full information estimator, which grows in size as more information is obtained.
Due to this, the solution $z^K_t$ of $\Scale[0.98]{\mathbb{P}_t(x^{\textup{prior}}_{t-M_t},{\mathbf{u}}_t,{\mathbf{y}}_t)}$ has a lower dimension compared to the solution $\Scale[0.969]{z^K_{t+1}}$ of $\Scale[0.969]{\mathbb{P}_{t+1}(x^{\textup{prior}}_{t-M_t + 1},{\mathbf{u}}_{t+1},{\mathbf{y}}_{t+1})}$ when $t < M$.
To ensure the warm-starting step can be smoothly carried out for time steps $t < M$, we use the matrix 
\begin{align} \label{lambda definition}
    \Sigma_t := \left\{\begin{array}{l}
\operatorname{blkdiag}(\mathbf{I}^{n_{z_t} - n_x - n_y},\mathbf{0}^{n_x+n_y}), \; t < M, \\
\mathbf{I}^{n_{z_t}}, \quad\quad\quad\quad\quad\quad\quad\quad\quad\quad\;\;\,\; t \geq M,
\end{array}\right.
\end{align}
in Step 2 to map $z^K_t$ to $\Sigma_t z^K_t$, which has the same dimension as $z^K_{t+1}$.
In Step 3, the operator $\Upsilon({\mathbf{y}}_{t}, y_t)$ appends $y_t$ to the end of the sequence ${\mathbf{y}}_{t}$ for all $t \geq 0$, and discards the first element $y_{t-M_t}$ in $\mathbf{y}_{t}$ if $t > M$.

\begin{algorithm}[t!]
\textbf{\textup{Require: }} $K$, $M$, $\Phi(\cdot)$, $z^0_{0}$, $x^{\textup{prior}}_0$, ${\mathbf{u}}_0$, ${\mathbf{y}}_0$; \\
\textbf{\textup{For}} $t = 0,1,2,\cdots$ \textbf{\textup{Do}} \\ 
 1. Obtain $\hat{x}^K_t$ by solving $\mathbb{P}_t(x^{\textup{prior}}_{t-M_t},{\mathbf{u}}_t,{\mathbf{y}}_t)$ for $K$ iterations using optimization algorithm $\Phi(\cdot)$ with initial solution $z^0_t$; \\
2. Warm-starting: $z^0_{t+1} \leftarrow \Sigma_t z^K_{t}$; \\
3. Update problem parameters: $x^{\textup{prior}}_{t-M_t} \leftarrow \hat{x}^K_{t-M_t \mid t}$, ${\mathbf{u}}_{t+1} \leftarrow \Upsilon({\mathbf{u}}_{t}, \pi(\hat{x}^K_t))$, ${\mathbf{y}}_{t+1} \leftarrow \Upsilon({\mathbf{y}}_{t}, y_t)$; \\
4. Apply $\pi(\hat{x}^K_t)$ to the system \eqref{re written system}; \\
\textbf{\textup{End}} 
\caption{Sub-Optimal MHE in Feedback Control}
\label{framework}                               
\end{algorithm}

\subsection{Interconnection of Three Subsystems}
We identify three dynamic subsystems from Alg. 1:
\begin{subequations} \label{inter system}
 \begin{align}
& \text{Subsys. 1:} \Scale[0.99]{\left\{\begin{array}{l}
x_{t+1} = A x_t + B \pi({x}_t + e_t)  + w^1_{t},\\
\,\,\, \,\,\,\,y_t = C x_t + w^2_{t},
\end{array}\right.} \label{sys 1}\\
& \text{Subsys. 2:} \Scale[0.99]{\,\,\,\,\,\,\,\,\, \epsilon_{t+1} = \Phi_{K}(\epsilon_t, x_{t},  {y}_{t},  {u}_{t}, e_t),} \label{sys 2}\\
& \text{Subsys. 3:} \Scale[0.99]{\,\,\,\,\,\,\,\,\, e_{t+1} = \mathcal{E}(e_t, x_t, \epsilon_t).} \label{sys 3}
\end{align}   
\end{subequations}
They describe the closed-loop controlled system (Subsys. 1), the sub-optimality error dynamics (Subsys. 2), and the estimation error dynamics (Subsys. 3), respectively.
Fig. 1 illustrates the interconnections between the three subsystems.

In subsystem 1, the controller $\pi(x_t)$ attempts to drive $x_t$ to the origin. However, $\pi(x_t)$ is perturbed to $\pi(\hat{x}^K_{t \mid t})$ by $e_t$.
In subsystem 2, $\mathbb{P}_t(x^{\textup{prior}}_{t-M_t},{\mathbf{u}}_t,{\mathbf{y}}_t)$ is solved for $K$ iterations with warm-starting to reduce the sub-optimality error (drive $z^0_{t} = \Sigma_{t-1} z^K_{t-1}$ to $z^*_{t}$). 
The optimal solution $z^*_{t}$ can be seen as a perturbed solution of $z^*_{t-1}$, resulting from the problem parameter update in Step 3 of Alg. 1.
In subsystem 3, the MHE attempts to drive the estimation error to zero. This process is disturbed by the change in state $x_t$ and the sub-optimality error $\epsilon_t$.
The stability of the interconnected system \eqref{inter system} can be analyzed via the small-gain theorem, which requires each subsystem to be ISS. Note that subsystem 1 in \eqref{sys 1} already meets this requirement via Assumption \ref{system ISS}.

\begin{figure}[t!]
 \centering
\includegraphics[width=0.7\hsize]{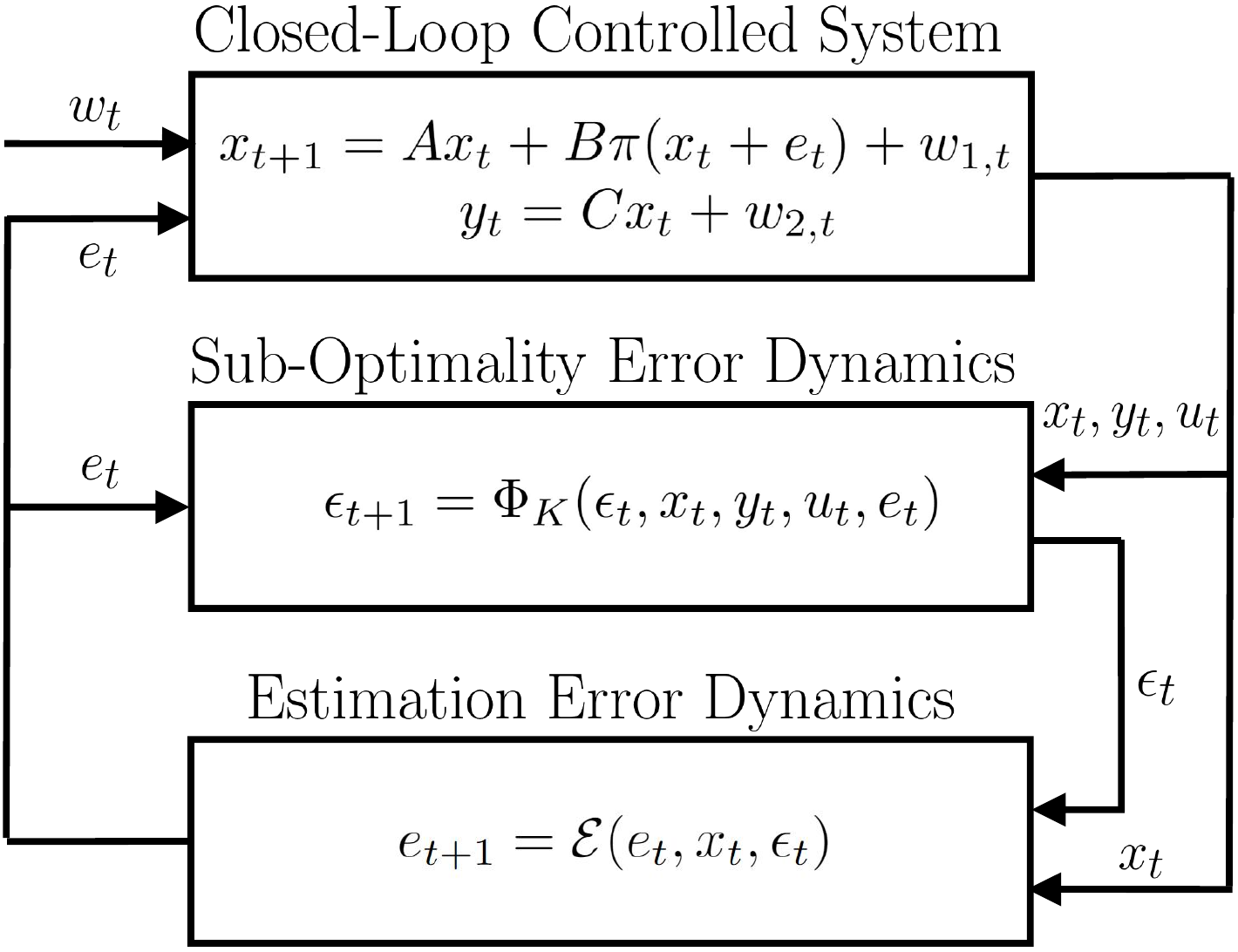}
\caption{The interconnection of three subsystems.}
\label{interco}
\end{figure}

\subsection{ISS of the Sub-Optimality Error Dynamics (Subsystem 2)}

To prove the sub-optimality error dynamics is ISS, we first show the difference between two consecutive optimal solutions $z^*_{t-1}$ and $z^*_{t}$ is bounded w.r.t. the changes in the parameters of $\mathbb{P}_t(x^{\textup{prior}}_{t-M_t},{\mathbf{u}}_t,{\mathbf{y}}_t)$. 

\begin{lemma}
    Suppose Assumptions 1-2 hold.
    Then, there exists a Lipschitz constant $L_{\Phi} > 1$ such that the optimal solutions of $\mathbb{P}_{t-1}(x^{\textup{prior}}_{t-M_t-1},{\mathbf{u}}_{t-1},{\mathbf{y}}_{t-1})$ and $\mathbb{P}_t(x^{\textup{prior}}_{t-M_t},{\mathbf{u}}_t,{\mathbf{y}}_t)$ satisfy 
    \begin{align}\label{MHE lipschitz 1}
       & \| \Sigma_{t-1} z^*_{t-1}  - z^*_{t} \| \leq   L_{\Phi}  (\| \Tilde{z}_{t-1} - \Sigma^\top_{t-1}\Tilde{z}_{t} \| + \sigma_{t}), 
    \end{align}
 with $\tilde{z}_t$ and $\tilde{z}_{t-1}$ defined in \eqref{compact definitions}, $\Sigma_{t}$ defined in \eqref{lambda definition}, and
    \begin{align} \label{sigma definition}
    \sigma_t := \left\{\begin{array}{l}
(1 - \eta^{-1})\| H_t\| + \| A \| + \| B \| + \| C \| + 2, \, t \leq M, \\
0, \quad\quad\quad\quad\quad\quad\quad\quad\quad\quad\quad\quad\quad\quad\quad\quad\;\,\, t > M.
\end{array}\right.
\end{align}

   \begin{proof} We prove \eqref{MHE lipschitz 1} by treating $\mathbb{P}_t(\cdot)$ as a parametric optimization problem, whose cost function is strongly convex (from Assumption 2), inequality constraints are convex, and equality constraints are affine. 
For $t > M$, using Theorem 3.1 in \cite{hager} and the fact $\Sigma_{t} = \mathbf{I}^{n_{z_t}}$ for $t \geq M$ from \eqref{lambda definition}, we know there exists a Lipschitz constant $L_{\Phi} > 1$ such that 
\begin{align} \label{lemma 1 inter}
    \|\Sigma_{t-1} z^*_{t-1}  - z^*_{t} \| \leq L_{\Phi} \| \Tilde{z}_{t-1} - \Sigma^\top_{t-1} \Tilde{z}_{t} \|.
\end{align}

For $t \leq M$, we consider an equivalent expression of $\mathbb{P}_t(\hat{x}_0,{\mathbf{u}}_{t},{\mathbf{y}}_{t})$, given by $\mathbb{P}_t'(\hat{x}_0, \allowbreak {\mathbf{u}}_{t}, \allowbreak {\mathbf{y}}_{t},H_{t}, \allowbreak 
 A, \allowbreak B, \allowbreak \mathbf{I}^{n_{x}}, C,\mathbf{I}^{n_{y} })$.
 The matrix $H_{t}$ is from the cost \eqref{MHE cost condensed}.
The last five matrices are from the system constraints \eqref{sys constraint 1} and \eqref{sys constraint 2}, $i \allowbreak = \allowbreak t-1$, respectively.
Let $\check{\mathbb{P}}_{t} := \mathbb{P}_t'(\hat{x}_0, \allowbreak {\mathbf{u}}_{t}, \allowbreak {\mathbf{y}}_{t}, \allowbreak \eta^{-1} H_{t},  \allowbreak \mathbf{0}^{n_{x}},  \allowbreak \mathbf{0}^{n_{x}\times n_u}, \allowbreak \mathbf{0}^{n_{x}}, \allowbreak \mathbf{0}^{n_{y}\times n_u}, \allowbreak \mathbf{0}^{n_{y}})$, with optimal solution $\check{z}^*_t$.
With $A, \allowbreak B, \allowbreak \mathbf{I}^{n_{x}}, C,\mathbf{I}^{n_{y}} =  \mathbf{0}$, $\check{\mathbb{P}}_{t}$ is equivalent to $\mathbb{P}_{t-1}(\hat{x}_0,{\mathbf{u}}_{t-1},{\mathbf{y}}_{t-1})$ with inactive system constraints at $i = t-1$. Thus, $\check{z}^*_t = \Sigma_{t-1} z^*_{t-1}$.
Similar to \eqref{lemma 1 inter}, we know there exists $L_{\Phi} > 1$ such that the optimal solutions of $\check{z}^*_t$ and $z^*_{t-1}$ satisfy
\begin{align} \label{lemma 2 inter}
    \|\check{z}^*_t - z^*_{t} \| \leq L_{\Phi} \sigma_t \Rightarrow  \|\Sigma_{t-1} z^*_{t-1}  - z^*_{t} \| \leq L_{\Phi} \sigma_t.
\end{align}

Since $\| \Tilde{z}_{t-1} - \Sigma^\top_{t-1}\Tilde{z}_{t} \| = 0$ for $t \leq M$ and $\sigma_t = 0$ for $t > M$, we can combine \eqref{lemma 1 inter} and \eqref{lemma 2 inter} to obtain \eqref{MHE lipschitz 1}.
   \end{proof}
\end{lemma}

With the bound in \eqref{MHE lipschitz 1}, we can show the sub-optimality error dynamics defined in \eqref{sys 2} is ISS:

\begin{proposition}
Consider $\mathbb{P}_t(x^{\textup{prior}}_{t-M_t},{\mathbf{u}}_t,{\mathbf{y}}_t)$ solved by an optimization algorithm $\Phi(\cdot)$ for $K$ iterations.
Suppose Assumptions 1-5 hold. For $t \geq 0$, the sub-optimality error $\epsilon_t$ satisfies
     \begin{align} \label{sys3 trajectory function}
          &   \| \epsilon_t \|  {\leq}   \beta_2(\| \epsilon_0\|,t) + \gamma_{2,1}( \| \mathbf{x}_{[0,t-1]} \| ) +  \gamma_{2,3}( \| \mathbf{e}_{[0,t-1]} \| )  \nonumber \\
           &\quad\quad\; + \gamma^{w}_{2}( \| \mathbf{w}_{[0,t-1]} \|)+ \gamma^{\sigma}_{2}(\| \bm{\sigma}_{[0,t-1]}\|), 
     \end{align}
where $\beta_2(s,t) := \phi(K)^t s$, 
$\gamma_{2,1}(s) := C_1(K)/{(1-\phi(K))} s$,
$\gamma_{2,3}(s) := C_2(K)/{(1-\phi(K))} s$,
$\gamma_{2}^w(s) := C_3(K)/{(1-\phi(K))} s$,
and $\gamma_{3}^\sigma(s) := \phi(K) L_{\Phi}/{(1-\phi(K))} s$,
with $C_1(K)$,
$C_2(K)$,
and $C_3(K)$ defined in \eqref{c1}-\eqref{c3}.
\end{proposition}

\subsection{ISS of the Estimation Error Dynamics (Subsystem 3)}
Inspired by \cite{muller_lyapu}, we first construct an $M$-step Lyapunov function for \eqref{sys 3} based on $W_\delta(\cdot)$ defined in \eqref{ioss function definition}. 

\begin{proposition}
    Suppose Assumptions 1-5 hold. 
Let $\bar{H} := \sup_{ t \geq 0}(\overline{\lambda}(H_t))$.
For $t \geq 0$, the state estimate $\hat{x}^K_{t \mid t}$ satisfies
\begin{align} \label{M step}
 &  W_\delta(\hat{x}^K_{t \mid t}, x_t)\leq  6 \eta^{M_t} W_\delta(\hat{x}^K_{t-M_t \mid t - M_t}, x_{t-M_t}) \nonumber \allowdisplaybreaks \\ 
& \quad\quad\quad +  2\bar{H}\| \epsilon_t \|^2 + 6 \sum_{j=1}^{M_t} \eta^{j-1} \|w_{t-j}\|^2_Q. 
\end{align}
\end{proposition}
 
Based on the $M$-step Lyapunov function in \eqref{M step}, we show the estimation error dynamics is ISS.

\begin{proposition}
Suppose Assumptions 1-5 hold.
Then, the estimation error dynamics is ISS and $\hat{x}^K_{t \mid t}$ satisfies
\begin{align} \label{sys2 trajectory function}
 \| e_t \| \leq  & \beta_3(\| e_0 \|,t) + \gamma_{3,1}(\| \mathbf{x}_{[0,t-1]} \|)  + \gamma_{3,2}(\|\bm{\epsilon}_{[0,t-1]} \|) \nonumber\\
 &   + \gamma^w_{3}(\|\mathbf{w}_{[0,t-1]}\|) + \gamma^{\sigma}_{3}(\| \bm{\sigma}_{[0,t-1]}\|),
\end{align}
where $\beta_3(s,t) :=  C_{e}(K) \sqrt{\rho}^{t} s$, 
$\gamma_{3,1}(s) := \sqrt{ 2 \Lambda^{\bar{H}}_P } C_1(K) s$,
$\gamma_{3,2}(s) := C_{\epsilon}(K) s$,
$\gamma_{3}^w(s) :=  C_{w}(K) s$,
and $\gamma_{3}^\sigma(s) :=\sqrt{ 2 \Lambda^{\bar{H}}_P }\phi(K)  L_{\Phi} s$, with $\rho$ satisfying $\rho^M = 6 \eta^M$ and $ C_{e}(K)$, $C_w(K)$, and $C_{\epsilon}(K)$ defined in \eqref{c4}-\eqref{c6}.
\end{proposition}

\subsection{Stability of the Interconnected System}

Given that subsystems 1, 2, and 3 are ISS satisfying \eqref{sys1 trajectory function}, \eqref{sys3 trajectory function}, and \eqref{sys2 trajectory function}, respectively, we can establish conditions on the iteration number $K$ such that the small-gain theorem is satisfied and the interconnected system is ISS.

\begin{theorem} \label{final theorem 1}
Consider the interconnected system \eqref{inter system}. Suppose Assumptions 1-5 hold.
Then, for any $K$ satisfying 
\begin{align} 
    & \gamma_{1,3}\circ \gamma_{3,1}(s) < s,  \label{sg 1}\\ 
    & \gamma_{2,3} \circ \gamma_{3,2}(s) < s,  \label{sg 2}\\
    & \gamma_{1,3}\circ \gamma_{3,2} \circ \gamma_{2,1}(s) < s, \label{sg 3}
\end{align}
for all $s > 0$, the interconnected system \eqref{inter system} is ISS w.r.t. the augmented disturbance $w_t$ and virtual disturbance $\sigma_t$.
\end{theorem}

\begin{remark}
    Since $\gamma_{1,2}, \gamma_{1,3}, \gamma_{2,1}, \gamma_{2,3}, \gamma_{3,1}, \gamma_{3,2} \in \mathcal{K}$, and
    $\gamma_{2,1}, \gamma_{2,3}, \gamma_{3,1} \rightarrow 0$ as $K \rightarrow \infty$,
    there always exists a iteration number ${K}$ such that \eqref{sg 1}-\eqref{sg 3} are satisfied.
\end{remark}


\section{Case study with an MHE-MPC}

\begin{figure}[t!]
 \centering
\includegraphics[width=1\hsize]{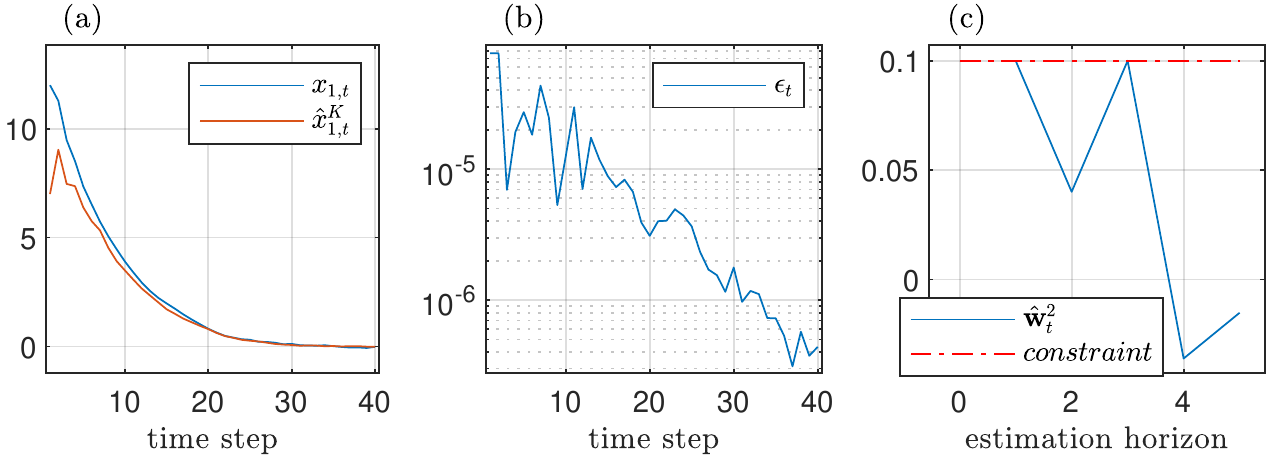}
\caption{(a) True state vs. Sub-optimal estimate; (b) Change in sub-optimality error; (c) The estimated measurement noise obtained from solving the MHE problem at $t = 6$.}
\label{interco}
\end{figure}

To demonstrate Alg. 1 and the theoretical findings, we consider the discrete-time linear system and the corresponding MPC controller in the case study of \cite{dominic1}.
We add an output matrix $C= [0.1, 0.3, 0.8, 0.5]$ to the system such that the system is observable.
The state $x \in \mathbb{R}^4$ and measurement $y \in \mathbb{R}$ are unconstrained, and the input $u \in [-1,1]\times[-1,1]$. 
Each element of the disturbance vector $w_t$ is sampled independently and uniformly from $[-0.1, 0.1]$.
We found $\gamma_{1,3}(s) := 28.8 s$, through the method used in Proposition 2 of \cite{our_journal}, and $L_{\pi} = 2.65$, through a sample-based method.

The parameters of the MHE problem in \eqref{MHE formulation} are $M = 5$, $Q = \textbf{I}^4$, $R = 1$, and $\eta = 0.8$, with $P$ computed to satisfy \eqref{ioss condition}.
Problem \eqref{MHE formulation} is written in a condensed form and solved using the partial gradient method \cite{dominic1} with convergence rate $\| z^K_{t} - z^*_{t} \| \leq 0.98^K \| z^{0}_{t} - z^*_{t} \|$.
Accordingly, we define $\phi(K) := 0.98^K$.
The Lipschitz constant $L_{\Phi} = 5.32$ is determined through a sample-based method.
Finally, the iteration number $K = 652$ is computed, which satisfies \eqref{sg 1}-\eqref{sg 3} with the previously defined parameters.

Given an initial state $x_0 =[12, -10, 10, -10]^\top$, $z^0_0 = x^{\textup{prior}}_0 = [7, -7, 3, -5]^\top$, and empty sequences $\textbf{y}_0$ and $\textbf{u}_0$, Alg. 1 is applied for 40 time steps.
Fig. 2(a) shows the state $x_{1,t}$ converges asymptotically to a neighbourhood of $0$ and the sub-optimal estimate $\hat{x}^K_{1,t}$ converges asymptotically to a neighbourhood of $x_{1,t}$.
Fig. 2(b) shows that the sub-optimality error $\epsilon_t$ converges asymptotically to a neighbourhood of $0$.
Thus, subsystems 1-3 defined in \eqref{sys 1}-\eqref{sys 3} are ISS.
Fig. 2(c) shows the estimated measurement noise sequence $\hat{\mathbf{w}}^{2,K}_t$ obtained from solving \eqref{MHE formulation} at time step $t = 6$, which respects the constraint (in red) by the design of MHE.

\section{Conclusion}
In this work, we proposed a sub-optimal MHE scheme applied to the control of linear systems with constraints. By characterizing Alg. 1 as three interconnected subsystems, we derived conditions on the optimization iteration number for guaranteeing ISS of the interconnected system w.r.t. to external disturbance and measurement noises. 
A possible extension is to consider the stability of systems controlled by sub-optimal MPC combined with sub-optimal MHE in applications with limited computation resources.

\section{Appendix}

We define some terms here for clarity:
\begin{align}
  &  C_1(K) := 2 \phi(K)L_{\Phi}(1+M (\| C\|+L_{\pi})) \label{c1} \allowdisplaybreaks \\
  &  C_2(K) :=  2\phi(K) L_{\Phi} (1+M  L_{\pi}) \label{c2} \allowdisplaybreaks\\
  &  C_3(K)  := 2\phi(K)  L_{\Phi} M \label{c3} \allowdisplaybreaks\\
& C_{e}(K) := 2 \sqrt{3 \Lambda^P_P \Lambda^{\bar{H}}_P } \phi(K) L_{\Phi} ( \sqrt{\rho}^{-M} +  L_{\pi} {\sqrt{\rho}^{-1}} ) \nonumber \allowdisplaybreaks \\
     & \quad\;\; + 4 \sqrt{3 \Lambda^P_P \Lambda^{\bar{H}}_P } \phi(K) L_{\Phi} L_{\pi} \sum^{M-1}_{i=1}  {\sqrt{\rho}^{-1-i}} + \sqrt{ 6 \Lambda^P_P } \nonumber \allowdisplaybreaks \\
      & \quad\;\;  + 2 \sqrt{3 \Lambda^P_P \Lambda^{\bar{H}}_P } \phi(K) L_{\Phi}  (L_\pi + 1) {\sqrt{\rho}^{-M-1}}, \label{c4} \allowdisplaybreaks\\
    & C_w(K) :=  \sqrt{ 2 \Lambda^{\bar{H}}_P } C_3(K) + \sqrt{ 6 \Lambda^Q_P } ({1-\sqrt{\rho}})^{-1} \nonumber \allowdisplaybreaks \\
    & \quad\;\; +  4 \sqrt{ 3 \Lambda^{\bar{H}}_P \Lambda^Q_P}\phi(K)   L_{\Phi} \left(  L_{\pi} M  + 1  \right) ({1-\sqrt{\rho}})^{-1}, \label{c5} \allowdisplaybreaks \\
   & C_{\epsilon}(K) :=  \sqrt{ 2 \Lambda^{\bar{H}}_P } \phi(K)  +  \sqrt{ 2 \Lambda^{\bar{H}}_P } ({1-\sqrt{\rho^M}})^{-1} \nonumber  \allowdisplaybreaks\\
& \quad\;\;+  4 \Lambda^{\bar{H}}_P \phi(K) L_{\Phi} \left(  L_{\pi} M  + 1  \right) ({1-\sqrt{\rho^M}})^{-1}. \label{c6}
\end{align}

\noindent \textbf{Proof of Proposition 1:} 
We break the proof into two cases.

\noindent \textbf{Case 1}: For $t \leq M$, due to the warm-start step (Step 2) in Alg. 1, we have 
     \begin{align}
           & \| z^0_t - z^*_t\|  = \| \Sigma_{t-1} z^K_{t-1} - z^*_t\| \\
            & \leq  \| \Sigma_{t-1} z^K_{t-1} - \Sigma_{t-1} z^*_{t-1}\| + \|  \Sigma_{t-1} z^*_{t-1} - z^*_t\| \\
            & \overset{\eqref{MHE lipschitz 1}}{\leq}  \| \Sigma_{t-1} \|\| z^K_{t-1} - z^*_{t-1}\| +  L_{\Phi} \sigma_t. \label{t less M}
        \end{align}
By multiplying $\phi(K)$ on both sides of the above inequality, and using \eqref{complexity bound} and the fact $\| \Sigma_{t} \| = 1 \forall t \in \mathbb{R}$, we have
\begin{align} 
    \| \epsilon_t \|  {\leq} \phi(K)  \| \epsilon_{t-1}\| + \phi(K)  L_{\Phi} \| \bm{\sigma}_{[0,t-1]}\|. \label{sys3 lyapunov function 1}
\end{align}
where we bounded $\sigma_t$ with $\| \bm{\sigma}_{[0,t-1]}\|$.

\noindent \textbf{Case 2}: For $t > M$, due to the warm-start step (Step 2) in Alg. 1, and $\Sigma^\top_{t} = \textbf{I}^{n_{z_t}}$ and $\sigma_{t} = 0$ for $t > M$, we have 
        \begin{align}
           & \| z^0_t - z^*_t\|  \leq  \| z^K_{t-1} - z^*_{t-1}\| + \| z^*_{t-1} - z^*_t\| \\
            &  \overset{\eqref{MHE lipschitz 1}}{\leq}   \| z^K_{t-1} - z^*_{t-1}\| + L_{\Phi} \| \Tilde{z}_{t-1} - \Tilde{z}_t\| \\
            & \leq L_{\Phi} \sum^{M_t-1}_{i=0} (\| {u}_{t-1-i } - {u}_{t - 2-i } \| 
 + \| {y}_{t-1-i } - {y}_{t -2-i } \| )\nonumber \\
            &+  L_{\Phi}  \| \hat{x}^K_{t-M \mid t-M} - \hat{x}^K_{t-M-1 \mid t-M-1} \| + \| \epsilon_{t-1}\|, \label{lemma 2 0}
        \end{align}
where we used $x^{\textup{prior}}_{t-M} = \hat{x}^K_{t-M \mid t-M}$ and $x^{\textup{prior}}_{t - M - 1} = \hat{x}^K_{t - M - 1 \mid t - M - 1}$ in \eqref{lemma 2 0}.
Given the above inequality,
we can bound $\| \hat{x}^K_{t-M \mid t-M} - \hat{x}^K_{t-M-1 \mid t-M-1} \|$ with
    \begin{align}
    & \| \hat{x}^K_{t-M \mid t-M} - \hat{x}^K_{t-M-1 \mid t-M-1} \| \nonumber \\
= & \| ( {x}_{t-M} + {e}_{t-M} ) - ( {x}_{t-M-1} + {e}_{t-M-1} ) \|  \\
\leq & \| {x}_{t-M_t} \| + \| {x}_{t-M_t-1}\| + \|{e}_{t-M_t} \| +  \| {e}_{t-M_t-1}\|, \label{lemma 2 1}
\end{align}
bound $ \| {u}_{t-1-i } - {u}_{t - 2-i } \|$ with 
 \begin{align}
 & \| {u}_{t-1-i } - {u}_{t - 2-i } \|
 \overset{\eqref{controller lip constant}}{\leq}  L_{\pi}  \| \hat{x}^K_{t-1-i} - \hat{x}^K_{t-2-i} \| \nonumber \\
& \Scale[0.985]{\leq  L_{\pi} (\| {x}_{t-1-i} \| + \| {x}_{t-2-i}\| + \|{e}_{t-1-i} \| + \| {e}_{t-2-i}\| ),} \label{lemma 2 2}
 \end{align} 
and bound $\| {y}_{t-1-i } - {y}_{t -2-i } \|$ with 
\begin{align}
  &  \| {y}_{t-1-i } - {y}_{t -2-i } \| \leq    \| {w}_{t-1-i } \| + \| {w}_{t-2-i }\|\nonumber \allowdisplaybreaks \\
  & + \| C\| \|{x}_{t-1-i } \| + \| C\|  \| {x}_{t -2-i } \|. \label{lemma 2 3}
\end{align}
Using the resulting bound to replace the term $\| z^0_t - z^*_t\|$ on the r.h.s. of \eqref{complexity bound}, we have that
\begin{align}
 &   \|\epsilon_{t} \| \leq \phi(K)  \| \epsilon_{t-1}\| +  C_1(K)  \| \mathbf{x}_{[0,t-1]} \|  + C_3(K) \| \mathbf{w}_{[0,t-1]} \| \nonumber \allowdisplaybreaks \\
 & +  \phi(K)   L_{\Phi}  \left( \|{e}_{t-M} \| +  \| {e}_{t-M-1}\|  \right) \nonumber \allowdisplaybreaks  \\
  & + \phi(K)   L_{\Phi} \sum^{M-1}_{i=0} \left( L_{\pi} ( \|{e}_{t-1-i} \| + \| {e}_{t-2-i}\| ) \right). \label{lemma 2 5}
\end{align}
where the $\| x_t\|$ and $\| w_t\|$ terms are bounded with $\| \mathbf{x}_{[0,t-1]} \|$ and $\| \mathbf{w}_{[0,t-1]} \|$, respectively.
Next, bounding the $\| e_t\|$ terms in \eqref{lemma 2 5} with $\| \mathbf{e}_{[0,t-1]} \|$ gives 
\begin{align}
  &  \| \epsilon_t \|  {\leq}  \phi(K)  \| \epsilon_{t-1}\| + C_1(K)  \| \mathbf{x}_{[0,t-1]} \|  \nonumber \\
  & \quad   +C_2(K)  \| \mathbf{e}_{[0,t-1]} \| + C_3(K) \| \mathbf{w}_{[0,t-1]} \|. \label{sys3 lyapunov function 2}
\end{align}
Combining the r.h.s. of \eqref{sys3 lyapunov function 1} and \eqref{sys3 lyapunov function 2} gives
    \begin{align} 
    & \| \epsilon_t \|  {\leq}  \phi(K)  \| \epsilon_{t-1}\| + C_1(K)  \| \mathbf{x}_{[0,t-1]} \| +  C_2(K)  \| \mathbf{e}_{[0,t-1]} \|  \nonumber \\
  & \quad    + C_3(K) \| \mathbf{w}_{[0,t-1]}\| +\phi(K) L_{\Phi} \| \bm{\sigma}_{[0,t-1]}\|, \label{sys3 lyapunov function}
    \end{align}
 which holds for all time steps $t > 0$. 
Finally, applying \eqref{sys3 lyapunov function} for $t$ times and using the geometric series to simplify $\sum_{i=0}^{t-1} \phi(K)^{(t-1-i)}$ as $1/(1-\phi(K))$ yield \eqref{sys3 trajectory function}.
\hfill \qedsymbol
\newline

\noindent \textbf{Proof of Proposition 2:} 
We first derive an intermediate bound on $W_\delta(\hat{x}^K_{t \mid t}, x_t)$.
 Due to Assumption \ref{optimizer condition}, the sub-optimal solution $(\hat{\mathbf{x}}^K_t, \hat{\mathbf{y}}^K_t, \hat{\mathbf{w}}^K_t )$ is feasible for \eqref{MHE formulation} and forms a feasible trajectory of the system in \eqref{controlled system}.
Given the actual trajectory $( {\mathbf{x}}_{[t-M,t]}, {\mathbf{y}}_{[t-M,t-1]},{\mathbf{w}}_{[t-M,t-1]} )$, we can apply the bound in \eqref{ioss function} for $M_t$ times to obtain 
\begin{align}
& \Scale[0.99]{ W_\delta(\hat{x}^K_{t \mid t}, x_t) \leq \eta^{M_t} W_\delta(\hat{x}^K_{t-M_t \mid t}, x_{t-M_t}) } \nonumber \allowdisplaybreaks \\
 & \Scale[0.99]{ + \sum_{j=1}^{M_t} \eta^{j-1} ( \|\hat{w}^K_{t-j \mid t} - w_{t-j}\|^2_Q+ \| \hat{y}^K_{t-j \mid t } - y_{t-j}\|^2_R ) } \allowdisplaybreaks  \\
 &\Scale[0.99]{ \leq    2 \eta^{M_t} \| \hat{x}^K_{t-M_t \mid t} - \hat{x}^K_{t-M_t \mid t - M_t}\|^2_P } \nonumber \allowdisplaybreaks  \\
 & \Scale[0.99]{ + 2 \eta^{M_t} \| \hat{x}^K_{t-M_t \mid t - M_t } - x_{t-M_t } \|^2_P + \sum_{j=1}^{M_t} \eta^{j-1}  2 \|w_{t-j}\|^2_Q } \nonumber \allowdisplaybreaks  \\
  &  \Scale[0.99]{ + \sum_{j=1}^{M_t} \eta^{j-1} ( \| \hat{y}^K_{t-j \mid t } - y_{t-j} \|^2_R  + 2 \|\hat{w}^K_{t-j \mid t}\|^2_Q)  } \label{lemma 1 bb} \allowdisplaybreaks \\
& \Scale[0.99]{  \leq  2 \eta^{M_t} W_\delta(\hat{x}^K_{t-M_t \mid t - M_t}, x_{t-M_t}) 
 + \sum_{j=1}^{M_t} \eta^{j-1}  2 \|w_{t-j}\|^2_Q   } \nonumber \allowdisplaybreaks  \\
& \Scale[0.99]{ + V_{\textup{MHE}}(\hat{x}^K_{t - M_t\mid t} , \hat{\mathbf{y}}^K_{t},  \hat{\mathbf{w}}^K_{t})  }\label{lemma 1 bound}
\end{align}
where \eqref{lemma 1 bb} is obtained by applying the triangle inequality to $W_\delta(\hat{x}^K_{t-M_t \mid t}, x_{t-M_t})$ and $\|\hat{w}^K_{t-j \mid t} - w_{t-j}\|^2_Q$.
Next, we derive a bound on $V_{\textup{MHE}}(\hat{x}^K_{t - M_t\mid t} , \hat{\mathbf{y}}^K_{t},  \hat{\mathbf{w}}^K_{t} )$. We know that
\begin{align}
  &  V_{\textup{MHE}}(\hat{x}^K_{t - M_t\mid t}, \hat{\mathbf{y}}^K_{t},  \hat{\mathbf{w}}^K_{t}) = \| z^K_t - \Tilde{z}_t \|^2_{H_t} \\
    & \leq 2\| z^K_t - z^*_t \|^2_{H_t} + 2\| z^*_t - \Tilde{z}_t \|^2_{H_t} \\
    & \leq 2\| z^K_t - z^*_t \|^2_{H_t}  + 2 V_{\textup{MHE}} ( \hat{x}^*_{t - M_t \mid t} , \hat{\mathbf{y}}^*_{t},  \hat{\mathbf{w}}^*_{t} ) \\
     & \leq \Scale[0.99]{2\| \epsilon_t \|^2_{H_t} + 2 V_{\textup{MHE}}({{x}}_{t-M_t},{\mathbf{y}}_{[t-M_t,t-1]}, {\mathbf{w}}_{[t-M_t,t-1]})} \label{prop 1 inter}
\end{align}
where \eqref{prop 1 inter} holds since $({\mathbf{x}}_{[t-M_t,t]},{\mathbf{y}}_{[t-M_t,t-1]}, {\mathbf{w}}_{[t-M_t,t-1]})$ forms a sub-optimal solution to \eqref{MHE formulation}.
Using the above bound with \eqref{lemma 1 bound} and then using \eqref{MHE cost} give
\begin{align} 
&  \Scale[0.99]{W_\delta(\hat{x}^K_{t \mid t}, x_t)\leq 2 \eta^{M_t} W_\delta(\hat{x}^K_{t-M_t \mid t-M_t}, x_{t-M_t})}  \nonumber \allowdisplaybreaks \\
&   \Scale[0.98]{ + \sum_{j=1}^{M_t} \eta^{j-1}  2 \|w_{t-j}\|^2_Q  +  2\| \epsilon_t \|^2_{H_t}} \nonumber \allowdisplaybreaks \\
& + 2 V_{\textup{MHE}}({x}_{t-M_t},{\mathbf{y}}_{[t-M_t,t-1]}, {\mathbf{w}}_{[t-M_t,t-1]}) \allowdisplaybreaks \\
& = 6 \eta^{M_t} W_\delta(\hat{x}^K_{t-M_t \mid t-M_t}, x_{t-M_t}) +  \sum_{j=1}^{M_t} \eta^{j-1}  6 \|w_{t-j}\|^2_Q \nonumber \allowdisplaybreaks \\ 
&  +  2\| \epsilon_t \|^2_{H_t}.
\end{align}
Lastly, using $\| \epsilon_t \|^2_{H_t} \leq \overline{\lambda}(H_t)\| \epsilon_t \|^2 \leq \bar{H}\| \epsilon_t \|^2$ in the last equality gives \eqref{M step}. \hfill \qedsymbol
\newline

\noindent \textbf{Proof of Proposition 3:} 
Let $t = c M + l$, with $l \in \mathbb{I}_{[0,M-1]}$ and $c \in \mathbb{I}_{\geq 0}$. At time step $l$, plugging $M_t = l$ into \eqref{M step} gives
%
%
%
\begin{align} \label{prop 2 1}
W_\delta (\hat{x}^K_{l \mid l}, x_l)\leq &  6 \eta^l W_\delta(\hat{x}^K_{0 \mid 0}, x_{0}) +  2 \bar{H} \| \epsilon_l \|^2 \nonumber \allowdisplaybreaks \\ 
 &  +  6 \sum_{j=1}^{l} \eta^{j-1} \|w_{l-j}\|^2_Q.
\end{align}

At time step $t$, applying \eqref{M step} for $c$ times, and bounding the resulting $W_\delta (\hat{x}^K_{l \mid l}, x_l)$ with \eqref{prop 2 1} gives 
\begin{align} 
& W_\delta(\hat{x}^K_{t \mid t}, x_t)  \leq  \rho^{kM} 6 \eta^l W_\delta(\hat{x}^K_{0 \mid 0}, x_{0}) + 2 \bar{H} \sum_{i = 0}^{c-1} \rho^{iM} \| \epsilon_{t-iM} \|^2  \nonumber \allowdisplaybreaks 
 \\ 
& + 2 \bar{H} \rho^{kM} \| \epsilon_{l} \|^2   + 6 \sum_{i = 0}^{c}  \rho^{iM}\sum_{j=1}^{M} \eta^{j-1} \|w_{t-iM-j}\|^2_Q  \label{prop 2 2} \allowdisplaybreaks  \\
& \leq 2 \bar{H} \sum_{i = 0}^{c} \rho^{iM} \| \epsilon_{t-iM} \|^2  + 6 \sum_{j = 0}^{t-1}  \rho^{j} \|w_{t-j-1}\|^2_Q    \nonumber \allowdisplaybreaks  \\ 
& + 6 \rho^{t} W_\delta(\hat{x}^K_{0}, x_{0})  \label{prop 2 3}
\end{align} 
where $\rho^M$ is used to replace $6\eta^M$ in \eqref{prop 2 2}. 
To obtain \eqref{prop 2 3}, $\rho$ is used to bound $\eta$, since $\rho/\eta = 6^{1/M} > 1$. 
Then, applying the bounds $\underline{\lambda}(P) \| e_t \|^2 \leq W_\delta(\hat{x}^K_{t \mid t}, x_t) \leq \overline{\lambda}(P) \| e_t \|^2$ and $\|w_{t}\|^2_Q \leq \overline{\lambda}(Q) \|w_{t}\|^2$ to \eqref{prop 2 3} gives
\begin{align}  \label{prop 2 10}
&   \| e_t \|^2 \leq   6 \Lambda_P^Q  \sum_{j = 0}^{t-1}  \rho^{j} \|w_{t-j-1}\|^2  + 2 \Lambda^{\bar{H}}_P  \sum_{i = 1}^{c} \rho^{iM} \| \epsilon_{t-iM} \|^2  \nonumber \\
& \quad  + 2 \Lambda^{\bar{H}}_P \| \epsilon_{t} \|^2 + 6 \rho^{t}  \Lambda_P^P  \| e_0 \|^2.
\end{align} 
Finally, by bounding $\|w_{t-j-1}\|$ with $\|\mathbf{w}_{[0,t-1]}\|$, bounding $\| \epsilon_{t-iM} \|$ with $\|\bm{\mathbf{\epsilon}}_{[0,t-1]}\|$,
taking square roots on both sides of \eqref{prop 2 10} using $\sqrt{a+b} \leq \sqrt{a} + \sqrt{b}$, and applying the geometric series, we obtain
\begin{align} \label{prop 2 4}
 & \| e_t \| \leq  \sqrt{ 6 \Lambda^P_P } \sqrt{\rho}^{t} \| e_0 \| + \sqrt{ 6 \Lambda^Q_P } (1-\sqrt{\rho})^{-1} \|\mathbf{w}_{[0,t-1]}\| \nonumber \\
 & +\sqrt{2 \Lambda^{\bar{H}}_P }  ({1-\sqrt{\rho^M}})^{-1} \|\bm{\epsilon}_{[0,t-1]} \|  +\sqrt{ 2 \Lambda^{\bar{H}}_P }     \|\epsilon_{t} \|.
\end{align}
To eliminate $\|\epsilon_{t}\|$ in \eqref{prop 2 4}, we consider two cases:

\noindent \textbf{Case 1}: For $t \leq M$, $\|\epsilon_{t} \|$ can be bounded by \eqref{sys3 lyapunov function 1} to obtain 
\begin{align} \label{prop 2 11}
 & \| e_t \| \leq  \sqrt{ 6 \Lambda^P_P } \sqrt{\rho}^{t} \| e_0 \| + \sqrt{ 6 \Lambda^Q_P } ({1-\sqrt{\rho}})^{-1} \|\mathbf{w}_{[0,t-1]}\| \nonumber \allowdisplaybreaks \\
 & +\sqrt{ 2 \Lambda^{\bar{H}}_P }  (({1-\sqrt{\rho^M}})^{-1}+\phi(K)) \|\bm{\epsilon}_{[0,t-1]} \|  \nonumber \allowdisplaybreaks \\
 & + \sqrt{2 \Lambda^{\bar{H}}_P } \phi(K)  L_{\Phi} \| \bm{\sigma}_{[0,t-1]}\|.  
\end{align}
where the resulting $\| \epsilon_{t-1}\|$ is bounded by $\| \bm{\epsilon}_{[0,t-1]]}\|$.

\noindent \textbf{Case 2}: For $t > M$, $\|\epsilon_{t} \|$ can be bounded by \eqref{lemma 2 5} to obtain 
\begin{align} 
 & \| e_t \| \leq  \sqrt{ 6 \Lambda^P_P } \sqrt{\rho}^{t} \| e_0 \|  +  \sqrt{ 2 \Lambda^{\bar{H}}_P } C_1(K)  \| \mathbf{x}_{[0,t-1]} \| \nonumber \allowdisplaybreaks \\
 & + (\sqrt{ 6 \Lambda^Q_P } ({1-\sqrt{\rho}})^{-1} + \sqrt{ 2 \Lambda^{\bar{H}}_P } C_3(K) ) \|\mathbf{w}_{[0,t-1]}\|  \nonumber \allowdisplaybreaks\\
 & +\sqrt{ 2 \Lambda^{\bar{H}}_P }  (({1-\sqrt{\rho^M}})^{-1}+\phi(K)) \|\bm{\epsilon}_{[0,t-1]} \|  \nonumber \allowdisplaybreaks \\
 &+ \sqrt{ 2 \Lambda^{\bar{H}}_P }\phi(K)   L_{\Phi} \sum^{M-1}_{i=0} \left( L_{\pi} ( \|{e}_{t-1-i} \| + \| {e}_{t-2-i}\| ) \right)  \nonumber \allowdisplaybreaks \\
  & +  \sqrt{ 2 \Lambda^{\bar{H}}_P } \phi(K)   L_{\Phi}  \left( \|{e}_{t-M} \|  +  \| {e}_{t-M-1}\|  \right).   \label{lemma 3 last}
\end{align}
Using \eqref{prop 2 4} to bound $\|{e}_{t-1-i} \|$, $\| {e}_{t-2-i}\|$, $i \in [0,M-1]$ in \eqref{lemma 3 last} and simplifying the expression gives 
\begin{align} \label{prop 2 12}
 & \| e_t \| \leq  C_{e}(K) \sqrt{\rho}^{t} \| e_0 \|  + \sqrt{ 2 \Lambda^{\bar{H}}_P } C_1(K)  \| \mathbf{x}_{[0,t-1]} \|  \nonumber \\
 & + C_{\epsilon}(K) \|\bm{\epsilon}_{[0,t-1]} \|  + C_{w}(K) \|\mathbf{w}_{[0,t-1]}\|.
\end{align}
Since $C_e(K) \geq \sqrt{ 6 \Lambda^P_P }$, $C_{\epsilon}(K) \geq \sqrt{ 2 \Lambda^{\bar{H}}_P }  (({1-\sqrt{\rho^M}})^{-1}+\phi(K))$, and $C_{w}(K) \geq \sqrt{ 6 \Lambda^Q_P } ({1-\sqrt{\rho}})^{-1}$, we can combine \eqref{prop 2 11} and \eqref{prop 2 12} to obtain \eqref{sys2 trajectory function}, which holds for $t \geq 0$. \hfill \qedsymbol

\bibliographystyle{IEEEtran}
\bibliography{IEEEabrv,MC}

\end{document}